\documentclass{Interspeech}

\interspeechcameraready

\usepackage{multirow}
\usepackage[table,xcdraw]{xcolor}
\newcommand{\verticalrow}[2]{\parbox[t]{2mm}{\multirow{#1}{*}{\rotatebox[origin=c]{90}{#2}}}}

\title{{RESOUND}: Speech {\underline{Re}}construction from {\underline{S}}ilent Videos via Ac{\underline{ou}}stic-Sema{\underline{n}}tic {\underline{D}}ecomposed Modeling}

\author[affiliation={1}]{Long-Khanh}{Pham}
\author[affiliation={1}]{Thanh V. T.}{Tran}
\author[affiliation={2}]{Minh-Tan}{Pham}
\author[affiliation={1}]{Van}{Nguyen}

\affiliation{FPT Software AI Center}{Hanoi}{Vietnam}
\affiliation{IRISA}{ Université Bretagne Sud, UMR 6074, Vannes}{France}
\email{{\{}khanhpl2, thanhtvt1, vannth19{\}}@fpt.com; minh-tan.pham@irisa.fr}
\keywords{lip-to-speech synthesis, source-filter theory, speech decomposition}

\usepackage{comment}

\begin{document}

\maketitle

\begin{abstract}    
    Lip-to-speech (L2S) synthesis, which reconstructs speech from visual cues, faces challenges in accuracy and naturalness due to limited supervision in capturing linguistic content, accents, and prosody. In this paper, we propose RESOUND, a novel L2S system that generates intelligible and expressive speech from silent talking face videos. Leveraging source-filter theory, our method involves two components: an acoustic path to predict prosody and a semantic path to extract linguistic features. This separation simplifies learning, allowing independent optimization of each representation. Additionally, we enhance performance by integrating speech units, a proven unsupervised speech representation technique, into waveform generation alongside mel-spectrograms. This allows RESOUND to synthesize  prosodic speech while preserving content and speaker identity. Experiments conducted on two standard L2S benchmarks confirm the effectiveness of the proposed method across various metrics.
\end{abstract}

\section{Introduction}\label{sec: Introduction}

The increasing demand for seamless human-computer interaction has driven significant interest in lip-to-speech (L2S) technology due to its broad range of real-world applications. L2S holds great potential across various domains, from enhancing communication in noisy environments to providing assistive technologies for individuals with aphonia. 

However, existing methods \cite{He_Zhao_Ren_Liu_Huai_Yuan_2022, kim2024let, kim2021lip,  mira2022end} have primarily been evaluated on small, controlled datasets \cite{harte2015tcd,cooke2006audio,9156430}. These datasets come with limited vocabularies and few speakers, making them unsuitable for real-world scenarios. While current models perform well under such conditions, their accuracy drops considerably when they encounter large-scale, real-world datasets that involve the challenges of diverse speaker characteristics, unrestricted vocabularies, and background noise. Furthermore, achieving a balance between natural prosody and precise speech generation in those settings remains a major challenge. To address the one-to-many mapping issue and enhance linguistic fidelity in video-to-speech generation, researchers have explored integrating semantic information through additional modalities.
Current approaches have successfully introduced new techniques to enhance speech reconstruction and semantic content, including self-supervised  learning representation \cite{kim2024let, hsu2023revise, choi23_interspeech, lei2024uni}, text labels \cite{kim2023lip}, and auxiliary models such as automatic speech recognition \cite{kim2023lip, yemini2023lipvoicer} and lip-to-text (L2T) systems \cite{hegde2023towards, yemini2023lipvoicer}. Furthermore, studies that integrate multiple strategies \cite{yemini2023lipvoicer, kim2023lip} often implement them inefficiently, limiting their combined potential. For instance, \cite{hegde2023towards} incorporates L2T models but lacks an effective fusion mechanism and sufficient linguistic supervision, relying solely on spectrogram loss.
Another research direction focuses on using advanced models and training algorithms to improve speech dynamics. Techniques such as flow-based methods \cite{kim2024let, He_Zhao_Ren_Liu_Huai_Yuan_2022} and GANs \cite{kim2021lip, mira2022end} yield detailed, noise-free audio waveforms, yet they often face accuracy issues like mispronunciations and omissions. Recently, diffusion-based systems \cite{10377743, yemini2023lipvoicer} have generated highly intelligible and natural speech on real-world datasets, though they typically suffer from slower inference speeds due to the extensive sampling steps required.

 The source-filter theory \cite{tokuda2021source, Fant_1960} models human speech production as a two-stage process, where the excitation signal from vocal fold vibrations determines pitch (fundamental frequency), and the vocal tract acts as a filter, shaping formant frequencies. FastPitchFormant \cite{bak21_interspeech} pioneered neural acoustic modeling with the source-filter theory, enabling intelligible speech synthesis across pitch shifts. Expanding on this, MultiVerse \cite{bak-etal-2024-multiverse} integrated this principle into zero-shot Text-to-Speech (TTS) models. 
 Unlike TTS, the L2S task must reconstruct not only speech expressiveness but also its content. Therefore, while the source-filter theory holds promise, its application in L2S demands deeper exploration and novel advancements.
 

In this paper, we propose RESOUND, a novel speech synthesis framework designed to reconstruct missing audio from silent video. Our approach offers two key contributions. First, we enhance multimodal fusion by introducing a more efficient mechanism that integrates prosody- and linguistics-aware generation. Prosody is estimated from a brief speaker prompt using pitch, energy, and timbre to capture speaker-specific characteristics, while linguistic information is derived from visual features and text predictions from an L2T model to improve speech intelligibility. Second, we introduce a dual-path architecture inspired by source-filter theory, which decouples acoustic and semantic modeling, enabling their independent learning. This separated structure leads to the improvement of speech accuracy and naturalness. To the best of our knowledge, this is the first study to incorporate source-filter theory into L2S synthesis.
Furthermore, RESOUND enhances speech expressiveness and intelligibility by generating an audible waveform through the incorporation of both mel-spectrogram and speech units.
Extensive experimental evaluations demonstrate that our method consistently outperforms baselines across multiple evaluation metrics, while ablation studies further validate its effectiveness. Audio samples and source code are available on demo page\footnote{\url{https://resound-l2s.github.io/}}.

\section{Methodology}\label{sec: Methodology}
\begin{figure*}
    \vspace{-0.4cm}
    \centering
    \includegraphics[width=\textwidth]{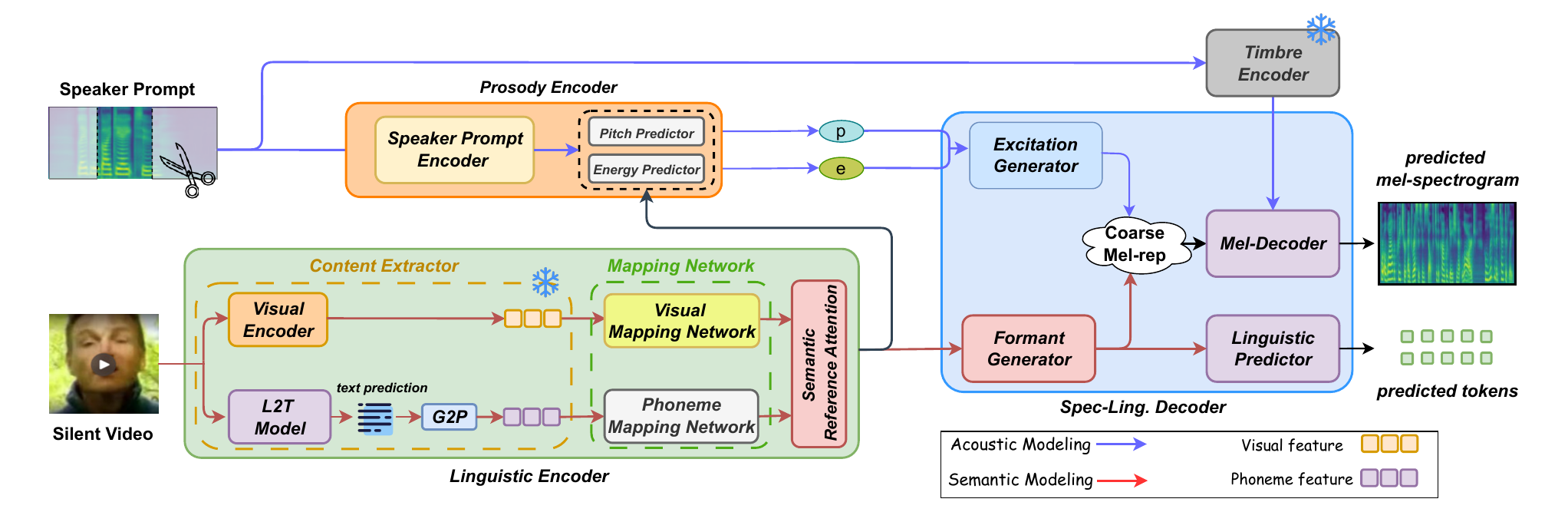}
    \caption{{RESOUND} framework:  RESOUND decomposes speech generation into acoustic and semantic branches. The acoustic stream extracts prosodic features (pitch, energy, timbre) via the Prosody and Timbre Encoders, while the semantic stream captures content from silent video using the Context Extractor, Mapping Network, and Semantic Reference Attention. The Spec-Ling Decoder then fuses both streams to generate a mel-spectrogram and speech tokens, which are processed by a vocoder for waveform synthesis.}
    \label{fig:OverallMethod}
    \vspace{-0.4cm}
\end{figure*}
Speech generation from a silent video in RESOUND is decomposed into two branches: an acoustic branch responsible for prosody modeling, and a semantic branch focused on linguistic representation. The framework consists of three core components: (1) Prosody Encoder, which extracts acoustic features conditioned on the speaker prompt; (2) Linguistic Encoder, which integrates phoneme and visual cues for linguistic modeling; and (3) Spec–Ling Decoder (Mel-spectrogram-Linguistic Decoder), which synthesizes mel-spectrograms and speech units from the combined acoustic and linguistic inputs. The overall model architecture is illustrated in Figure \ref{fig:OverallMethod}.

\subsection{Prosody Encoder} \label{sec:ProsodyEncoder}
A key challenge in L2S synthesis is reconstructing speech expressiveness, particularly natural intonation, stress, and rhythm. While lip movements convey linguistic content, prosodic features must be inferred from supplementary cues.
Therefore, our approach employs an acoustic branch that models source representations containing prosodic information with minimal dependence on linguistic content. This approach is grounded in the source-filter theory and consists of two key components: the Timbre Encoder and the Prosody Encoder. 
The process begins with the extraction of the reference mel-spectrogram $R_{\text{mel}}$ from the input audio prompt. To maintain timbre consistency, the H/ASP model \cite{heo2020clovabaselinevoxcelebspeaker} is employed as the Timbre Encoder, which preserves speaker-specific spectral characteristics by supplying timbre features from the reference mel  $R_{\text{mel}}$  for subsequent mel-spectrogram generation.

Concurrently, the Speaker Prompt Encoder extracts prosody-related features  $E_\text{spk}$  from  $R_{\text{mel}}$. To effectively model speaker-dependent prosodic patterns, a multi-head attention mechanism aligns  $E_\text{spk}$ with the linguistic representations $E_\text{ling}$ produced by the Linguistic Encoder (detailed in Section \ref{sec:LinguisticEncoder}). This alignment provides a crucial representation $E_\text{spk\_ling}$  for subsequent prosodic modeling stages, ensuring a natural and speaker-consistent synthesis. For accurate frame-level pitch and energy modeling, we adopt the predictor architecture of FastSpeech 2 \cite{ren2022fastspeech2fasthighquality}, utilizing a Pitch Predictor to estimate the fundamental frequency ($F_0$) and an Energy Predictor to model speech intensity from $E_\text{spk\_ling}$. This framework ensures that the synthesized speech maintains natural expressiveness and clarity.

\subsection{Linguistic Encoder} \label{sec:LinguisticEncoder}


In contrast to prosody, which governs speech rhythm, linguistic encoding ensures content accuracy. To capture this distinction, we introduce the semantic branch, a fundamental component of the established theory, responsible for generating vocal tract filter-related representations with minimal dependency on prosodic features. To facilitate this process, we propose the Linguistic Encoder, which effectively integrates multimodal representations from silent videos. This encoder consists of three core components: \textbf{(i)} the Content Extractor, \textbf{(ii)}  the Inter-Modal Mapping Network, and \textbf{(iii)} the Semantic Reference Attention.

\subsubsection{Content Extractor}  \label{subsec:ContentExtractor}

For a given silent video $V$, we employ AVHuBERT \cite{shi2022learningaudiovisualspeechrepresentation}, a self-supervised learning model specifically designed to capture fine-grained visual-linguistic features. AVHuBERT extracts frame-wise linguistic representations as  $ E_v = \mathrm{AVHuBERT}(V)$ where $E_v \in \mathbb R^{T_v \times D_v}$,  \( T_v \) and \( D_v \)  represent the number of video frames and the dimension of video features, respectively.
In parallel, we integrate an advanced L2T model \cite{10096889} to generate textual predictions. To further enhance linguistic accuracy, we employ a Grapheme-to-Phoneme (G2P) model \cite{lee2020learningpronunciationforeignlanguage} that converts these textual outputs into phoneme embeddings, represented as $E_p = \mathrm{G2P} \big( \mathrm{L2T}(V) \big) \in \mathbb{R}^{T_p \times D_p}$. Here, \( T_p \) and \( D_p \) denote the number of phonemes and the hidden dimension of phoneme embeddings, respectively.
By integrating both visual-speech representations and phonetic representations, this framework enhances accurate and comprehensive modeling of linguistic content in silent videos.

\subsubsection{Inter-Modal Mapping Network}
To ensure the generated phoneme embeddings $E_p$ remain contextually synchronized and aligned with the speaker's lip movements, we process the visual features $E_v$ through a Visual Mapping Network $\mathcal F_v$. This network, composed of an MLP layer followed by a Transformer Encoder, further refines the extracted visual representations. Similarly, the Phoneme Mapping Network $\mathcal F_p$ incorporates a Phoneme Encoder \cite{ren2022fastspeech2fasthighquality} alongside an additional MLP layer to enhance phoneme representations. The corresponding transformations are formulated as follows:
\begin{align}
    H_v &= \mathcal F_v(E_v +  PE_{1:T_v}) \in \mathbb{R}^{T_v \times D_v} \\
   H_p &= \mathcal F_p(E_p +   PE_{1:T_p}) \in \mathbb{R}^{T_p \times D_p}
\end{align}
where $H_v$ and  $H_p$ represent the refined visual and phoneme features, respectively. $PE$ denotes the positional embedding, which inserts sequential information into representations.
\subsubsection{Semantic Reference Attention}
To improve the fusion and enrichment of shared latent space representations between visual features $H_v$ and phoneme embeddings $H_p$, we introduce the Semantic Reference Attention (SRA). This module integrates these representations via a two-step attention-based process. First, the Reference Transformer aligns visual features with phoneme embeddings, ensuring that silent lip movements are correctly associated with their corresponding phonetic segments. Second, the Semantic Attention mechanism, implemented using Conformer \cite{gulati20_interspeech}, refines feature interactions by effectively modeling multi-scale dependencies.

Specifically, the Reference Transformer iteratively refines the mapping of visual inputs to phoneme sequences, thereby reinforcing the phoneme-visual relationship as follows:
\[
    R_{V \to P}^{[i]} =
    \begin{cases}
        H_v & \text{if } i = 0, \\
        \mathrm{MultiHeadAttn} \big( R_{V \to P}^{[i-1]}, H_{p} \big) & \text{otherwise},
    \end{cases}
\]
where $R_{V \to P}^{[i]}$ represents the refined mapping of visual features to phoneme sequences at the $i$-layer. This refinement process is carried out using a multi-head attention mechanism, denoted as $\mathrm{MultiHeadAttn} (\cdot)$. In this mechanism, $R_{V \to P}^{[i-1]}$ acts as the query, while $H_p$ serves as both the key and the value.

To enhance multi-scale information exchange between modalities, we apply Conformer \cite{gulati20_interspeech} to the final mapping representation, $R^{[n]}_{V \to P}$, where $n$ is the number of Reference Transformer's layers. Unlike standard transformers, Conformer incorporates convolutional layers that refine local feature extraction while maintaining long-range dependencies. This enables Conformer to effectively capture both fine-grained articulatory details and broader phonetic structures, making it particularly well-suited for enriching speech representation.

\subsection{Spec-Ling. Decoder}
In line with the RESOUND motivation theory, the Spec-Ling Decoder employs a dual-pathway structure, transforming encoder outputs into two distinct hidden-state representations:
\begin{itemize}
    \item \textbf{Excitation Representation:} Generated by the Excitation Generator, this representation encodes glottal excitation signals (source information).
    \item \textbf{Formant Representation:} Produced by the Formant Generator, this representation captures vocal tract resonances (filter information).
\end{itemize}
 
\noindent These two representations are combined using a simple additive fusion method to form a coarse mel-spectrogram, which is then used as input to the Mel-Decoder for high-fidelity output generation. Furthermore, to ensure linguistic precision, the Spec-Ling Decoder predicts an additional representation in the form of discrete speech tokens. This complementary output, produced by the Linguistic Predictor, reinforces linguistic structure at a higher abstraction level and further refines the synthesized speech for both semantic and phonetic accuracy. Finally, these features are passed to a vocoder, which converts them into an audible waveform.

\subsection{Training Objective}
The total loss for RESOUND is expressed as:
\begin{equation}
\label{eq:loss_function}
    \mathcal{L}_{tot} = \lambda_{m} \mathcal{L}_{m} + \lambda_{p} \mathcal{L}_{p} + \lambda_{e} \mathcal{L}_{e} + \lambda_{u} \mathcal{L}_{u},
\end{equation}
\noindent where $\mathcal{L}_{m}$, $\mathcal{L}_{p}$, $\mathcal{L}_{e}$, and $\mathcal{L}_{u}$ represent the mel, pitch, energy, and speech unit losses, respectively, with weighting factors $\lambda_{m}=100$, $\lambda_{p}=0.1$, $\lambda_{e}=0.1$, and $\lambda_{u}=0.01$. To compute the prediction errors for mel ($\mathcal{L}_{m}$), pitch ($\mathcal{L}_{p}$), and energy ($\mathcal{L}_{e}$), we follow \cite{kim2024let} to employ the L1 reconstruction loss. Additionally, Additionally, the speech unit loss $\mathcal L_u$ is calculated via a cross-entropy loss with a label smoothing parameter $\alpha=0.1$.

\section{Experimental Setup}
\label{sec: Experimental-Setup}

\begin{table*}[h]
    \centering
    \footnotesize
    \caption{Performance comparison of different methods on LRS3-TED and LRS2-BBC datasets.}\label{tab:main}
    \resizebox{\textwidth}{!}{
        \begin{tabular}{c|lcccccccc}
            \toprule
            \multirow{2}{*}{Dataset} & \multirow{2}{*}{Method} & {\textit{Naturalness}} & {\textit{Speaker Sim}} & \textit{Content} & \multicolumn{2}{c}{\textit{Prosodic Discrepancy}} & \multicolumn{2}{c}{\textit{Low-level}} & \textit{Efficiency} \\
            & & \textbf{UTMOS}$\uparrow$ & \textbf{SECS}$\uparrow$ & \textbf{WER}$\downarrow$ & \textbf{MAE$_{\text{RMSE}}$}$\downarrow$ & \textbf{MAE$_{\boldsymbol{F_0}}$}$\downarrow$ & \textbf{ESTOI}$\uparrow$ & \textbf{MCD-DTW-SL}$\downarrow$ & \textbf{RTF}$\downarrow$ \\

            \midrule

            \verticalrow{7}{LRS3-TED} 
            & Ground-truth & 3.519 & 1.0 & 0.99 & 0.0 & 0.0 & 1.0 & 0.0 & - \\
            & Vocoder$^\dagger$ & 3.430 & 0.890 & 1.21 & 0.009 & 16.88 & 0.852 & 7.24 & - \\
            
            \cmidrule{2-10}
            
            & Multi-task \cite{kim2023lip} & 1.274 & 0.529 & 57.00 & \underline{0.031} & 91.33 & 0.277 & 10.55 &  0.133 \\
            & DiffV2S \cite{10377743} & \underline{2.910} & 0.627 & 36.62 & 0.034 & 64.86 & 0.283 & 11.14 & - \\
            & Intelligible L2S \cite{choi23_interspeech} & 2.680 & \underline{0.750} & 27.69 & 0.032 & \underline{55.42} & \underline{0.396} & \underline{9.64} & \textbf{0.041} \\
            & LipVoicer \cite{yemini2023lipvoicer} & 2.385 & 0.591 & \underline{21.04} & 0.035 & 81.63 & 0.212 & 11.21 & 14.117 \\
            & \textbf{RESOUND} 
            & \textbf{3.002}   
            & \textbf{0.777}   
            & \textbf{20.06}  
            & \textbf{0.025} 
            & \textbf{49.52} 
            & \textbf{0.423} 
            & \textbf{8.71} 
            &  \underline{0.063} \\
            
            \midrule
            
            \verticalrow{7}{LRS2-BBC}
            & Ground Truth & 3.016 & 1.0 & 1.46 & 0.0 & 0.0 & 1.0 & 0.0 & - \\
            & Vocoder$^\dagger$ & 2.836 & 0.841 & 2.15 & 0.043 & 19.92 & 0.810 & 10.28 & - \\
                             
            \cmidrule{2-10}
            
            & Multi-task \cite{kim2023lip} & 1.296 & 0.552 & 51.86 & 0.056 & 88.093 & 0.341 & 13.73 & 0.133 \\
            & DiffV2S \cite{10377743} & \underline{2.169} & 0.582 & 49.55 & 0.046 & 69.19 & 0.282 & 14.24 & - \\
            & Intelligible L2S \cite{choi23_interspeech} & 1.962 & \underline{0.708} & 36.75 & \underline{0.031} & \underline{59.11} & \underline{0.390} & \underline{11.36} &  \textbf{0.041}\\
            & LipVoicer \cite{yemini2023lipvoicer} &  2.088 & 0.575 & \textbf{17.04} & 0.039 & 85.71 & 0.250 & 11.49 & 14.117 \\
           & \textbf{RESOUND} 
            & \textbf{2.382}   
            & \textbf{0.730}   
            & \underline{28.55}  
            & \textbf{0.025} 
            & \textbf{51.79} 
            & \textbf{0.420} 
            & \textbf{10.56} 
            & \underline{0.063} \\
            \bottomrule
            \multicolumn{10}{l}{$\dagger$ indicates that the multi-input vocoder is provided with both the ground-truth mel-spectrogram and the ground-truth speech units as inputs.}  \\
        \end{tabular}
    }
\end{table*}

\subsection{Datasets}



 \textbf{LRS2-BBC} \cite{afouras2018deep} is an English audio-visual dataset from BBC programs, comprising over 220 hours of video. We use its pre-training and training sets for model development and the test set for inference.

\noindent\textbf{LRS3-TED} \cite{afouras2018lrs3tedlargescaledatasetvisual} is an English audio-visual dataset from TED/TEDx talks, featuring diverse speakers and a vocabulary of 50,000+ words across 438 hours of video. Following \cite{schoburgcarrillodemira22_interspeech}, we train on ~131,000 and test on around 1,300 utterances.

\subsection{Evaluation Metrics}

Following \cite{10377743}, we evaluate speech synthesis using UTMOS \cite{saeki22c_interspeech} for naturalness, SECS for speaker similarity, and WER for intelligibility. SECS scores are computed via \texttt{Resemblyzer}\footnote{\url{https://github.com/resemble-ai/Resemblyzer}}, while WER is derived from Auto-AVSR \cite{10096889}.
We further introduce MAE$_{F_0}$ and MAE$_{\text{RMSE}}$ to quantify prosodic discrepancies and RTF to measure system speed in generating one second of speech. Finally, we assess intelligibility using ESTOI \cite{10.1109/TASLP.2016.2585878} and acoustic distance via MCD-DTW-SL \cite{Chen_2022_CVPR}\footnote{Unlike prior work, we omit LSE-* metrics as some methods outperform the ground truth, compromising synchronization reliability.}.

\subsection{Implementation Details}
For visual, we use MediaPipe \cite{lugaresi2019mediapipeframeworkbuildingperception} to crop the mouth ROI $(96 \times 96)$ and apply random cropping $(88 \times 88)$, flipping, erasing, and temporal masking. For audio, speech is resampled to 16kHz, mel-spectrograms are extracted, and HuBERT-BASE with 200 K-means clusters is used for speech unit extraction. Following \cite{kim2023lip}, we randomly extract a 0.5s segment from the ground-truth audio to serve as an audio prompt.Additionally, to generate ground-truth data for the pitch and energy predictors, we extract $F_0$ from the speech signal using a Praat-based extractor\footnote{\url{https://github.com/YannickJadoul/Parselmouth}}. For energy, we calculate the frame-wise magnitude of the linear spectrogram.


The Pitch and Energy Predictors follow the FastSpeech 2 architecture \cite{ren2022fastspeech2fasthighquality} , while the Speaker Prompt module shares a similar design. We implement the Conformer from \cite{choi23_interspeech} with eight layers. Both branch generators use three FFT blocks (hidden dim: 512, attention heads: 8), and the Mel-Decoder is a 1D CNN. The model is trained for 80 epochs (40k steps per dataset) with an initial learning rate of \(10^{-3}\). As AV-HuBERT is pretrained on LRS3, we freeze it for LRS3 experiments and fine-tune it for LRS2. For waveform generation, we use the multi-input vocoder from \cite{choi23_interspeech}.

\section{Experimental Results}

\subsection{Quality Comparison}
Table~\ref{tab:main} presents a comprehensive evaluation of RESOUND on the LRS2 and LRS3 datasets, where our method consistently outperforms previous L2S models across multiple metrics. By leveraging source-filter theory to disentangle the speech generation process, RESOUND learns distinct representations that contribute to its superior overall performance. Specifically, the proposed network achieves a UTMOS of 3.002 (with the ground truth at 3.519) and an SECS of 0.777 (compared to 0.89 for a Vocoder approach). In terms of content accuracy, RESOUND records a significantly lower WER of 20.06, underscoring the effectiveness of its semantic branch in capturing linguistic nuances. Furthermore, the inclusion of specialized predictors reduces pitch and energy errors, while the dedicated acoustic branch enhances low-level acoustic quality. On the LRS2 dataset, our method also consistently outperforms previous SOTA approaches across multiple metrics - except for WER. This discrepancy can be explained by the unpredictable noise present in LRS2, which hinders accurate speech reconstruction and leads to lower performance compared to LipVoicer - a denoising diffusion model. Although LipVoicer demonstrates strength in content reconstruction, it tends to produce speech that lacks naturalness and perceptual quality, as reflected in reduced acoustic metrics like UTMOS, SECS, and MAE$_{F_0}$. Moreover, LipVoicer's inference speed is over 200 times slower than that of RESOUND, further emphasizing its potential for real-time applications.

To further assess speech quality, we conduct subjective Mean Opinion Score (MOS) tests on the LRS3-TED dataset. In this evaluation, 30 participants rate the quality of 20 randomly synthesized audio samples per method on a 5-point scale with 0.5-point increments. Table \ref{tab:mos} presents the MOS results along with their 95\% confidence intervals. These results align with the findings from the LRS2 and LRS3 experiments in Table \ref{tab:main}, indicating that RESOUND consistently generates intelligible speech that closely resembles speech in both naturalness and similarity.
\begin{table}[ht]
    \centering
    \small
    \setlength{\tabcolsep}{3pt}
    \caption{MOS results on LRS3-TED dataset.\label{tab:mos}}
    \begin{tabular}{lccc}
         \toprule
         Method & Naturalness$\uparrow$ & Intelligibility$\uparrow$ & Similarity$\uparrow$ \\
         \midrule
         Ground Truth & $4.80 \pm 0.05$ & $4.76 \pm 0.07$ & $4.92 \pm 0.04$ \\
         \midrule
         DiffV2S \cite{10377743} & $3.28 \pm 0.16$ & $2.98 \pm 0.18$ & $2.96 \pm 0.15$ \\
         Intelligible L2S \cite{choi23_interspeech} & $2.99 \pm 0.15$ & $3.36 \pm 0.15$ & $3.07 \pm 0.14$ \\
         LipVoicer \cite{yemini2023lipvoicer} & $2.64 \pm 0.16$ & $2.95 \pm 0.18$ & $2.40 \pm 0.15$ \\
         \textbf{RESOUND} & $\bf{3.86 \pm 0.13}$ & $\bf{3.86 \pm 0.15}$ & $\bf{3.57 \pm 0.14}$ \\
         \bottomrule
    \end{tabular}    
\end{table}
\begin{table}[ht]
    \centering
    \small 
    \setlength{\tabcolsep}{3pt} 
    \caption{Ablation study results on LRS3-TED dataset.}
    \label{tab:ablation}
    \resizebox{\columnwidth}{!}{ 
        \begin{tabular}{lccccc}
            \toprule
            Method & UTMOS$\uparrow$ & WER$\downarrow$ & SECS$\uparrow$ & MAE$_{F_0}$$\downarrow$ & ESTOI$\uparrow$ \\ 
            \midrule
            RESOUND & \textbf{3.002} & \textbf{20.06} & \textbf{0.777} & \textbf{49.52} & \textbf{0.423} \\
            $-$ w/o L2T \& SRA & 2.840 & 27.49 & 0.774 & 51.796 & 0.405 \\
            $-$ w/o Acoustic branch & 2.176 & 21.93 & 0.713 & 69.219 & 0.385 \\
            $-$ w/o Energy Predictor & 2.698 & 20.22 & 0.744 & 51.22 & 0.400 \\
            \bottomrule
        \end{tabular}
    }
\end{table}  
\subsection{Ablation Studies}
We conduct ablation studies of several key components of RESOUND on the LRS3 dataset to analyze their individual effects. 

\noindent\textbf{Effects of L2T \& SRA:}  As shown in Table~\ref{tab:ablation}, removing these content-aware modules leads to significant degradation in WER (a relative increase by 37\%) while speech quality remains largely unaffected. This further supports our hypothesis on the disentangling effects of the linguistic and acoustic branches.

\noindent\textbf{Effects of acoustic branch:}  To evaluate the acoustic branch, we remove both the Prosody Encoder and Excitation Generator. This results in a significant decline in speech quality metrics, most notably with naturalness and pitch error with relative degradation of 28\% and 40\%, respectively, underscoring the acoustic branch’s critical role and its independent learning of acoustic attributes.

\noindent\textbf{Effects of Energy Predictor:}  Finally, the fourth row in Table \ref{tab:ablation} shows that incorporating both pitch and energy—unlike \cite{bak21_interspeech}, which overlooks energy—results in higher-quality speech.

\section{Conclusion}
This paper presents RESOUND, a novel framework for reconstructing speech from silent videos. Grounded in source-filter theory, RESOUND separates speech generation into acoustic and semantic branches, capturing prosody and linguistic content. It integrates multiple modalities, utilizing pitch and energy for acoustics and text predictions for semantic. Experiments on LRS2 and LRS3 demonstrate its SOTA performance across multiple metrics. Future research could explore its adaptability to diverse languages and accents.

\clearpage

\bibliographystyle{IEEEtran}
\bibliography{mybib}

\end{document}